# Enhanced spin-orbit coupling in a heavy metal via molecular coupling

S. Alotibi[1], B.J. Hickey[1], G. Teobaldi[2,3,4,5], M. Ali[1], J. Barker[1], E. Poli[2], D.D. O'Regan[6,7], Q. Ramasse[1,8], G. Burnell,[1] J. Patchett,[9] C. Ciccarelli,[9] M. Alyami,[1] T. Moorsom[1] and O. Cespedes[1*]

[1]*School of Physics and Astronomy, University of Leeds, Leeds LS2 9JT, United Kingdom.*
[2]*Scientific Computing Department, Science and Technology Facilities Council, Didcot OX11 0QX, United Kingdom*
[3]*Beijing Computational Science Research Center, 100193 Beijing, China*
[4]*Stephenson Institute for Renewable Energy, Department of Chemistry, University of Liverpool, L69 3BX Liverpool, United Kingdom*
[5]*School of Chemistry, University of Southampton, Highfield, SO17 1BJ Southampton, United Kingdom*
[6]*School of Physics, Trinity College Dublin, The University of Dublin, Dublin 2, Ireland*
[7]*Centre for Research on Adaptive Nanostructures and Nanodevices (CRANN) and the SFI Advanced Materials and Bio-Engineering Research Centre (AMBER), Dublin 2, Ireland*
[8]*SuperSTEM, SciTech Daresbury Science and Innovation Campus, Keckwick Lane, Daresbury WA4 4AD, United Kingdom*
[9]*Cavendish Laboratory, University of Cambridge, JJ Thomson Avenue, Cambridge CB3 0HE, United Kingdom*

5d metals are used in electronic architectures because of their high spin-orbit coupling (SOC) leading to efficient spin ↔ electric conversion and strong magnetic interactions. When $C_{60}$ is grown on a metal, the electronic structure is altered due to hybridisation and charge transfer. The spin Hall magnetoresistance for Pt/$C_{60}$ and Ta/$C_{60}$ at room temperature are up to a factor 6 higher than for the pristine metals, with the spin Hall angle increased by 20-60%. At low fields of 1-30 mT, there is an anisotropic magnetoresistance, increased up to 700% at room temperature by $C_{60}$. This is correlated with non-collinear Density Functional Theory simulations showing changes in the acquired magnetic moment of transport electrons via SOC. Given the dielectric properties of molecules, this opens the possibility of gating the effective SOC of metals, with applications for spin transfer torque memories and pure spin current dynamic circuits.

---

[*] o.cespedes@leeds.ac.uk

The spin-orbit interaction is perhaps the most crucial mechanism in the design of magnetic structures and metal device physics. It determines the magnetocrystalline anisotropy, is key to the propagation and electrical conversion of spin currents, determines the magnitude of interfacial mechanisms such as the Dzyaloshinskii–Moriya interaction and guides new paths of research, such as the generation of Majorana fermions and energy band engineering of topological insulators[1-5]. The SOC also controls the efficiency of spin - charge conversion in the spin Hall, spin torque and spin Seebeck effects. All of these are key to reducing the power consumption and energy dissipation of computing and electronic devices, an issue that is quickly coming to the forefront of technology. However, currently we can only tune the SOC by static means, such as doping, preventing the design of architectures where spin, charge and magnetic interactions can be reversibly modified to enhance device performance or to acquire new functionalities.

The Spin Hall magnetoresistance (SHMR) can quantify the SOC in thin (~nm) heavy metal layers deposited on a magnetic insulator such as the yttrium iron garnet $Y_3Fe_5O_{12}$ (YIG) [6-8]. When an electric current $J_c$ flows in the metal, the spin Hall effect (SHE) induces a perpendicular spin current $J_s$, with the spin polarization $s$ parallel to the film surface. If the YIG magnetization $M$ is parallel to $s$, $J_s$ cannot flow into the magnet and a spin accumulation forms. The resistance is the same as a bare Pt wire. When $M$ is not parallel to $s$, the transverse component exerts a torque on the YIG magnetic moments, injecting spin current into the magnet. This opens a channel of dissipation for the spin current, reducing the inverse SHE contribution to $J_c$ so that the resistance of Pt appears to have increased [5,9]. The largest dissipation takes place when $M$ is perpendicular to $s$ and the maximum SHMR should occur. The SHMR is measured by rotating the angle β in Fig. 1a, with the applied $H$ field (and therefore $M$) always orthogonal to the electrical current, but varying from in-plane to out-of-plane, and therefore from parallel ($R_{min}$) to perpendicular to the spin polarization ($R_{max}$) [5,10,11]. The ratio of the spin to charge current is known as the spin Hall angle: $\theta_{SH} = |J_s|/|J_c|$ [12-14]. $\theta_{SH}$ has technological relevance, as it is correlated with the torque exerted on ferromagnets in spin transfer torque memories [15]. A larger SHMR and therefore increased $\theta_{SH}$ can result in lower power or smaller switching currents for such devices. By tuning the SOC in conventional magnetic insulator/metal structures with a molecular layer, we can also differentiate spin transport effects based on their physical origins [16,17].

At metallo-molecular interfaces, the electronic and magnetic properties of both materials change due to charge transfer and hybridisation [18-21]. This can lead to the emergence of spin ordering and spin filtering [22-25], or change the magnetic anisotropy [21,26,27]. Even though composed of light carbon, fullerenes with large curvature can produce a large spin-electrical conversion [2,28-30]. Here, we study the effect of metal/$C_{60}$ interfaces on the SHMR and anisotropic magnetoresistance (AMR) of YIG/Pt and YIG/Ta. We aim to: investigate the mechanisms behind spin orbit scattering at hybrid metal/$C_{60}$ interfaces, maximise technologically-relevant parameters, and open new paths of research towards tunable SOC.

Using shadow mask deposition, we grew two metal wires simultaneously on the same YIG substrate and, without breaking vacuum, covered one wire with 50 nm of $C_{60}$ –modifying the density of states (DOS) and transport properties of the metal. According to our density functional theory (DFT) calculations for Pt/$C_{60}$, 0.18-0.24 electrons per $C_{60}$ molecule are transferred [31], and the first molecular layer is metallised. This reduces the electron surface scattering, improving the residual resistance ratio (RRR) –Figs. 1b and SI [31]. Our Ta wires have a resistivity (~1-2 µΩ·m) and a negative temperature coefficient (~-500·10$^{-6}$ K$^{-1}$), consistent with a sputtered β-Ta phase [32]. Opposite to Pt, C$_{60}$ increases the resistivity of Ta –see Fig. 1b.

The change in resistivity as the magnetic field is rotated is fitted to a $\cos^2(\beta)$ function, and the amplitude is taken to be the SHMR [31]. The SHMR saturates once the applied field saturates the magnetisation out-of-plane, at 0.1-0.15 T for a YIG film 170 nm thick at 290 K, and no higher than 0.5 T for any measured condition. However, above this field range, other contributions such as Koehler MR, localisation and the Hanle effect can result in significant linear and parabolic contributions to the MR that would artificially enhance the SHMR ratio and θ$_{SH}$ (Fig 1c)[31,33]. For a YIG/Pt(2nm) sample, the C$_{60}$ layer increases the MR due to spin accumulation by about a factor 3, but reduces the polynomial contributions because of the increased effective (conducting) thickness of the Pt/C$_{60}$ bilayer. In YIG/Ta(4nm), where C$_{60}$ increases the resistance rather than reducing it, both the SHMR up to 0.15 T and the polynomial MR at higher fields are enhanced (Fig. 1d).

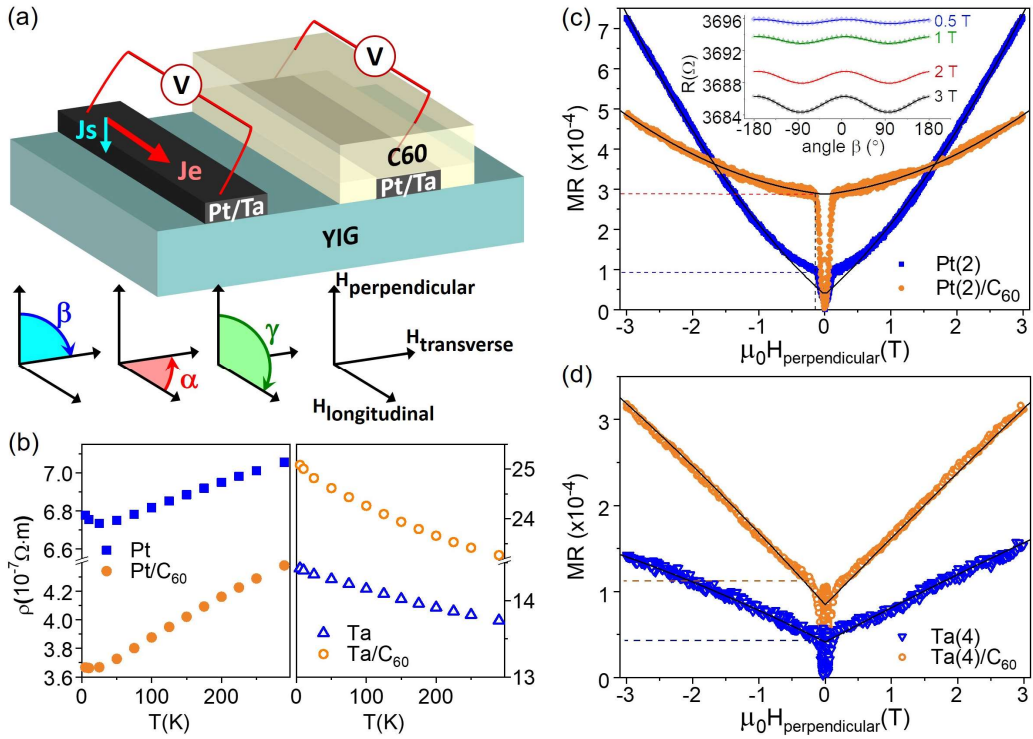

**FIG. 1** (a) Schematic of the experiment. There are three possible orientations of the magnetic field (H) w.r.t. the electrical current and the YIG film. To measure only the SHMR without AMR effects, we rotate H from perpendicular to transverse (change in β). (b) Typical resistivity of thin Pt (3 nm) and Ta (4 nm) wires on YIG. With $C_{60}$ on top, the Pt resistivity is about 40% lower, the RRR factor increases and the upturn at low T is absent. With Ta, we observe the opposite effect, an increase in the resistivity with the molecular interface. Inset: Resistance with different applied fields as a function of the angle β. The data is fitted to a $\cos^2(\beta)$ function. We take the amplitude at the lowest field of 0.5 T, when the YIG substrate is saturated but the polynomial contributions are small, as the SHMR value. (c) MR in a Pt wire with $H_{perpendicular}$. The spin Hall contribution to the MR at 300 K reaches a maximum at ~0.1-0.15 T, where the YIG film is saturated out of plane. (d) MR in a Ta wire with $H_{perpendicular}$ at 75 K. The maximum in the spin Hall contribution at this temperature is reached at ~0.2 T.

The SHMR values at 0.5 T for Pt and Pt/$C_{60}$ are plotted in Fig. 2a, and the ratios with and without a molecular overlayer in Fig. 2b. The temperature dependence of the SHMR reproduces observations in RF-sputtered YIG/sputtered Pt wires [33]. For Pt grown by evaporation on thicker, liquid epitaxy or pulsed laser deposition YIG, the SHMR has a gentler drop at high temperatures. This is attributed to a smaller temperature dependence of the spin diffusion length [34,35], which could be due to a different resistivity of Pt and different magnetic behaviour of YIG films depending on the growth method. It is possible that the larger SHMR observed in metallo-molecular wires could be due to a change in the spin mixing conductance ($G^{\uparrow\downarrow}$) induced by $C_{60}$ [6,36]. However, $G^{\uparrow\downarrow}$ is related to the spin transparency of the YIG/Pt interface, where the effect of the molecular interface is small [31]. Also, we do not observe an increase in the ferromagnetic resonant damping $\alpha$, proportional to $G^{\uparrow\downarrow}$, of YIG/Pt with $C_{60}$ interfaces (Fig. 2c) [31,37]. Furthermore, the SHMR versus temperature results cannot be fitted by changing $G^{\uparrow\downarrow}$ without also changing $\Theta_{SH}$ [31]. Fig. 2d shows the $\Theta_{SH}$ values taking $G^{\uparrow\downarrow}=4\times10^{14}\,\Omega^{-1}m^{-2}$ [8]; see the SI for other fitting values [34,38]. For Pt wires of ≤5 nm, there is an increase in $\Theta_{SH}$ with $C_{60}$. This effect disappears for thick wires (> 10nm), where the molecular interface does not significantly change the spin Hall angle. A similar molecular enhancement of the SHMR and $\Theta_{SH}$ is observed for Ta wires [31]. Molecules may affect the Rashba effect and spin texture of the metal, leading to changes in the effective SOC of the hybrid wire [39-41]. In our simulations, we consider the perpendicular dipole formed due to charge transfer at the Pt/$C_{60}$ interface and its associated potential step breaking symmetry [42]. However, this dipole is maximum at 2.5 nm, where the experiments show a local minimum. Our calculations point rather towards a mechanism mediated by the magnetic moment acquired by the transport electrons, resulting in spin-dependent charge flow [31].

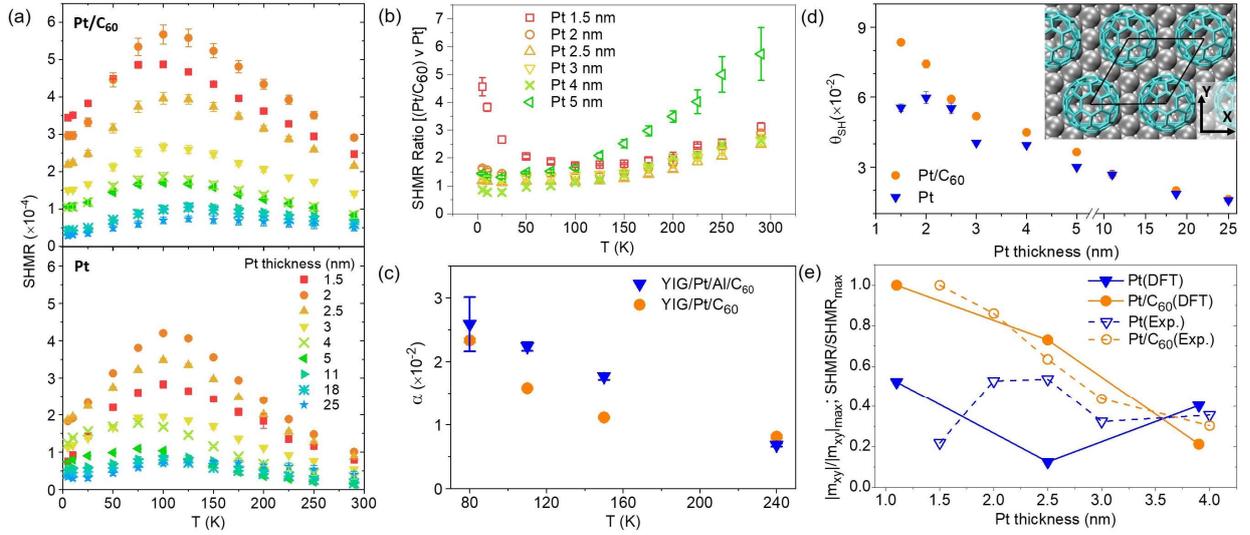

**FIG. 2** (a) SHMR for Pt and equivalent Pt/$C_{60}$ wires of different thicknesses on GGG/YIG(170 nm) films. (b) SHMR Ratios between Pt/$C_{60}$ and Pt. The maximum effect of the molecular layer (factor 4 to 7 change) take place for thin films (1.5 nm) at low temperatures or thick films (5 nm) at room temperature. (c) The magnetic resonance damping $\alpha$ is not increased by the $C_{60}$ interface; here a comparison of YIG/Pt/Al/$C_{60}$ and YIG/Pt/$C_{60}$ shows similar or even higher damping values for the decoupled Pt/Al/$C_{60}$ sample. (d) For wires ≤5 nm, $\theta_{SH}$ obtained from the SHMR data fits is significantly higher with the molecular overlayer. Inset: Top view of the optimized $C_{60}$/Pt(111)-(2√3x2√3)R30° interface DFT model. The $C_{60}$ molecules are adsorbed on top of one Pt-vacancy. The black polygon marks the in-plane periodicity of the system. Pt: silver, C: cyan. (e) DFT simulations of the electrical current-induced, in-plane magnetic moments ($|m_{xy}|$) and experimental SHMR, normalized to the largest calculated ($|m_{xy}|$) or measured value (SHMR) as a function of the Pt thickness.

Non-collinear band structure calculations enable analysis of the atom Projected (energy-dependent) Magnetization Density (PMD) for different Pt and Pt/$C_{60}$ film thicknesses. In all cases, we find the PMD for the in-plane (x,y) magnetic moment components ($m_{x,y}$) to be larger than for the out of plane one ($m_z$). It is also possible to observe an enhancement of the PMD oscillation magnitudes due to the adsorption of $C_{60}$. The effect becomes smaller as the Pt thickness increases from 1.1 nm to 2.5 nm and 3.9 nm, correlated with the SHMR values in Pt/$C_{60}$ (Fig. 2e). The differences in PMD between the $C_{60}$/Pt and Pt systems document the role of the Pt/$C_{60}$ interfacial re-hybridization, and ensuing changes in the electronic structure, for enhancing SOC-related anisotropies and spin transport in Pt-based systems.

The fabrication of YIG films can lead to elemental diffusion and defects that change the magnetic properties of the ferrimagnet and the interpretation of transport measurements [43]. Figs. 3a-b show atomic-resolution aberration corrected cross-sectional scanning transmission electron microscopy (STEM) images and electron energy loss spectroscopy (EELS) chemical maps. It is possible to observe, in addition to a certain level of surface roughness of the YIG film, an area close to the YIG surface and below the sputtered Pt wire into which some Pt metal may have diffused and formed a low density of nm-sized clusters (see also Fig. S4 in [31]). This diffusion can affect the magnetization and anisotropy direction at the surface of the YIG layer, originating the minor loops we observe in the perpendicular field direction in some YIG films [31,43].

For Pt grown on YIG, an additional change in resistance is observed at low magnetic fields <5-20 mT when the direction of an applied magnetic field is changed with respect to the electrical current. The origin of this AMR is controversial. It has been attributed to a proximity-induced magnetization of Pt, which is close to the Stoner criterion, but it is also claimed that there is no evidence for this induced magnetization [16,17]. The same effect is also seen in YIG/Ta. This low field AMR (LF-AMR) is characterized by the presence of peaks, positive or negative depending on the field direction, resembling the AMR observed in magnetic films with domain wall scattering [44,45]. Due to the SOC, in most magnetic materials domain walls reduce the resistance for in-plane fields, and increase it for out of plane fields. This domain wall AMR peaks at the coercive field $H_c$ of the magnet, for the greatest magnetic disorder and domain wall density. In YIG/Pt, the position of out-of-plane LF-AMR peaks coincides with the coercivity of the perpendicular minor YIG loops (Fig. 3c and [31]), which could point to a YIG surface layer with an out-of-plane easy axis. We find that the LF-AMR has the same shape and peak position with or without a molecular overlayer. However, the magnitude of the LF-AMR is larger when $C_{60}$ is present. This molecular effect is stronger for the perpendicular configuration (Fig. 3d), which may be due a larger perpendicular magnetic anisotropy induced by $C_{60}$, as reported for Co [21]. A larger LF-AMR is also observed in YIG/Ta when $C_{60}$ is deposited on top [40]. For YIG films grown on YAG substrates, the in-plane coercivity is increased by 1-2 orders of magnitude, and the LF-AMR peaks appear at higher fields, supporting the correlation between the AMR in Pt and the surface YIG magnetisation (Figs. S5-S7 in [31]).

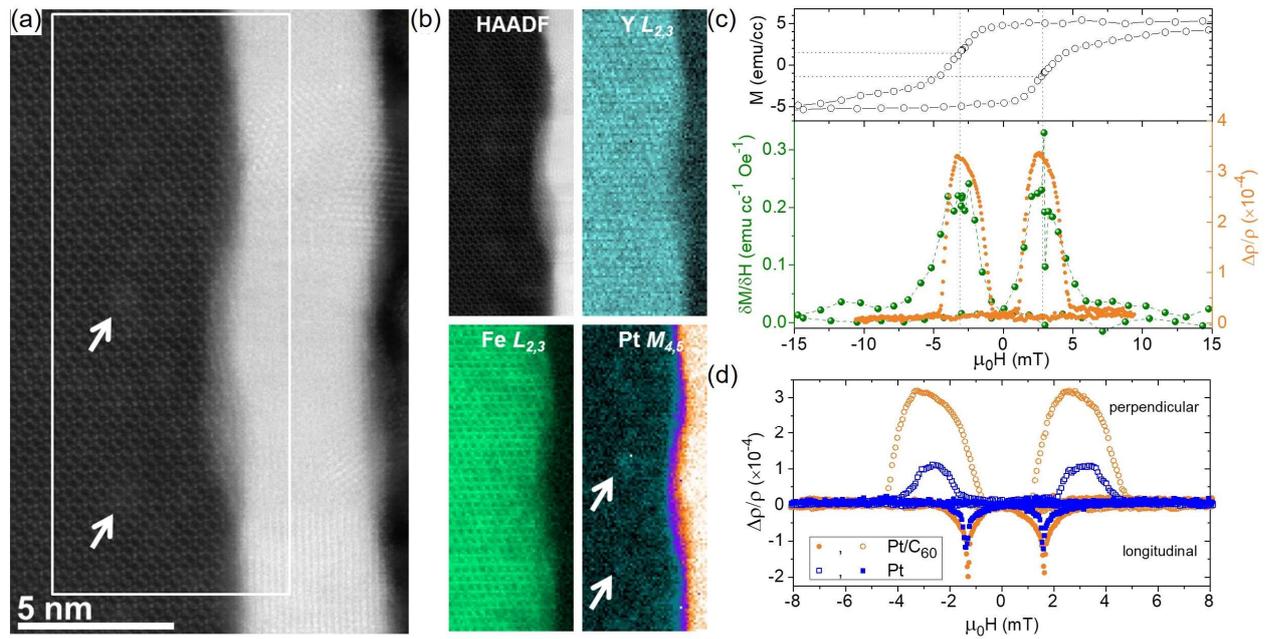

**FIG. 3** (a) Cross-sectional high angle annular dark field (HAADF) image of the YIG/Pt interface obtained using a scanning transmission electron microscope (see methods for details). (b) Elemental chemical analysis of the interface using EELS: the relative intensity maps of the Y, Fe and Pt ionization edges are presented with a simultaneously acquired HAADF image of the region, indicated by a white box in (a). Bright clusters immediately below the YIG surface, indicated by white arrows in the Pt map and the overview HAADF image, contain a higher Pt concentration and may be due to Pt diffusion into the YIG. (c) Low field MR and minor hysteresis loop with the field in the perpendicular orientation at 200 K. The full loop uncorrected and other examples can be found in [31]. (d) Room temperature LF-AMR comparison between YIG/Pt and YIG/Pt/$C_{60}$. The curves are qualitatively the same, but the magnitude of the effect is enhanced by the molecules.

The LF-AMR peak position (coercivity of the YIG surface) and peak width (saturation field of the YIG surface), increase as the temperature is lowered (Figs. 4a-b). Typically, the AMR of YIG/Pt measured at high fields is reported to vanish above 100-150 K. If measuring at 3 T, where quantum localisation and other effects are strong, we observe this same decay with temperature. However, the LF-AMR can be observed up to room temperature. $C_{60}$ not only increases the LF-AMR value, but it also makes it less temperature dependent, so that the LF-AMR ratio can be up to 700% higher for Pt/$C_{60}$ at 290 K. This supports our suggestion from DFT simulations of a mechanism based on $C_{60}$-induced re-hybridization enhancing the magnetic moment acquired by transport electrons via SOC (Fig. 4c).

The LF-AMR depends on the Pt thickness, $t$, as $(t-x)^{-1}$ (Fig. 4d). We identify the value of $x$, approximately 1 nm, as the magnetised Pt region contributing to the AMR. This relationship is not affected by the $C_{60}$ layer, although the magnitude is uniformly higher with molecules.

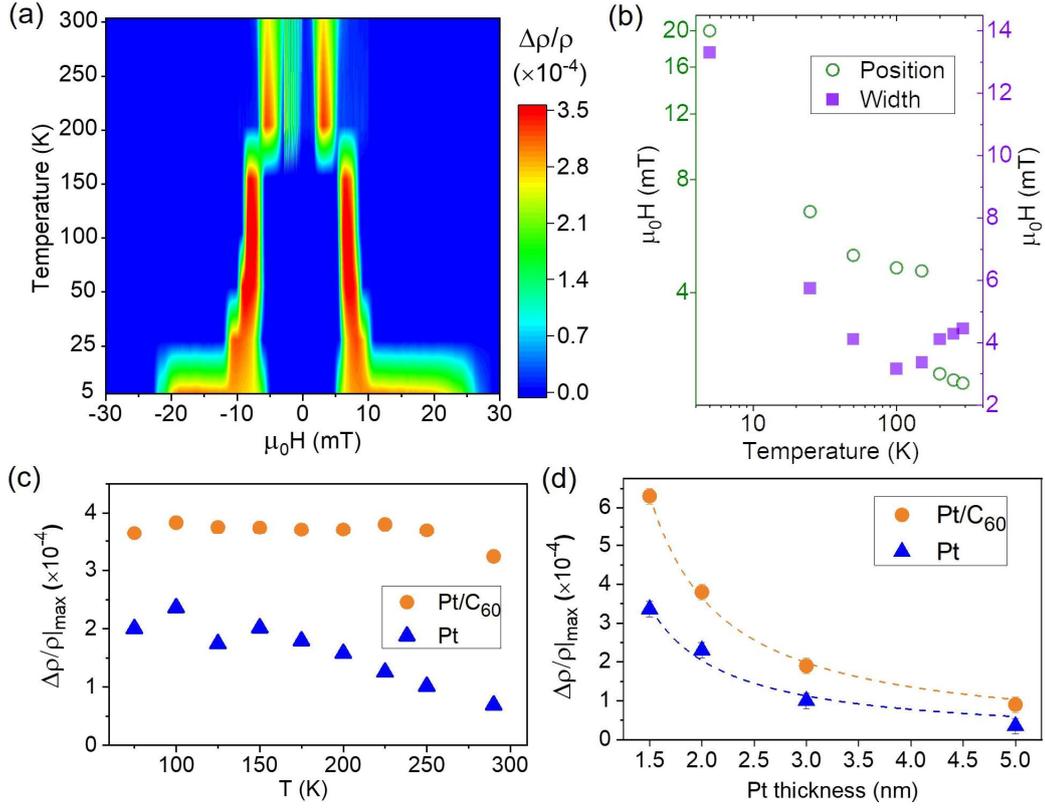

**FIG. 4** (a) Perpendicular LF-AMR for GGG/YIG(170)/Pt(2)/$C_{60}$(50). (b) As the sample is cooled, the perpendicular LF-AMR peak position and width are increased in steps, rather than monotonic fashion. (c) Temperature dependence of the maximum LF-AMR, calculated as the change in resistance from the peak in the perpendicular orientation to the minima in the longitudinal. There is a faster temperature drop in the MR values for Pt when compared with Pt/$C_{60}$. This may be due to the acquired magnetic moment in Pt/$C_{60}$ leading to a more stable induced magnetisation up to higher temperatures. (d) The LF-AMR for Pt and Pt/$C_{60}$ can be fitted to a $(t-x)^{-1}$ function, where $t$ is the Pt wire thickness and x is a constant of 1 nm that we identify with the magnetically active Pt region.

Our results show that molecular overlayers can enhance the spin orbit coupling of heavy metals, as observed in SHMR and AMR measurements. Additionally, the molecular layers aid in distinguishing the origin of spin scattering mechanisms, such as the coupling with YIG surface magnetisation and a LF-AMR measurable at high temperatures. The enhancement of the effective SOC with molecular interfaces has a wide range of applications, e.g. to reduce the current densities in spin transfer torque memories. Given the dependence on surface hybridisation and charge transfer, the effect could be controlled via an applied electrical potential. This is an important development, as nearly all other methods to alter the spin-orbit coupling of a material are static. The inverse SHE can be modified by gating with ionic liquids, but changes to the SOC are undetermined and the electrical conversion may only be quenched [46]. Materials can be doped during fabrication to increase the spin-orbit effect, but that becomes fixed in a circuit, i.e. static. Using UHV grown nanoscale molecular films that can be gated offers a dynamic response – the transport properties of an active circuit, e.g. to control the direction and magnitude of pure spin currents.

**ACKNOWLEDGMENTS**

This work was supported by Science Foundation Ireland [19/EPSRC/3605] and the Engineering and Physical Sciences Research Council (EPSRC) UK through grants EP/S030263, EP/K036408, EP/M000923, EP/I004483 and EP/S031081. This work made use of the ARCHER (via the UKCP Consortium, EPSRC UK EP/P022189/1 and EP/P022189/2), UK Materials and Molecular Modelling Hub (EPSRC UK EP/P020194/1) and STFC Scientific Computing Department's SCARF High-Performance Computing facilities. J.B. acknowledges support from the Royal Society through a University Research Fellowship. Electron microscopy work was carried out at SuperSTEM, the National Research Facility for Advanced Electron Microscopy supported by EPSRC. S.A. acknowledges support from Prince Sattam bin Abdulaziz University.

[1]	F. Hellman *et al.*, Reviews of Modern Physics **89**, Unsp 025006 (2017).
[2]	L. E. Hueso *et al.*, Nature **445**, 410 (2007).
[3]	J. Liu, F. C. Zhang, and K. T. Law, Physical Review B **88**, 064509 (2013).
[4]	W. J. Shi, J. W. Liu, Y. Xu, S. J. Xiong, J. Wu, and W. H. Duan, Physical Review B **92**, 205118 (2015).
[5]	J. Sinova, S. O. Valenzuela, J. Wunderlich, C. H. Back, and T. Jungwirth, Reviews of Modern Physics **87**, 1213 (2015).
[6]	Y. T. Chen, S. Takahashi, H. Nakayama, M. Althammer, S. T. B. Goennenwein, E. Saitoh, and G. E. W. Bauer, Physical Review B **87**, 144411, 144411 (2013).
[7]	H. Nakayama *et al.*, Physical Review Letters **110**, 206601, 206601 (2013).
[8]	M. Althammer *et al.*, Physical Review B **87**, 224401, 224401 (2013).
[9]	C. Hahn, G. de Loubens, O. Klein, M. Viret, V. V. Naletov, and J. Ben Youssef, Physical Review B **87**, 174417, 174417 (2013).
[10]	M. Althammer, Journal of Physics D-Applied Physics **51**, 313001 (2018).
[11]	J. Fontcuberta, H. B. Vasili, J. Gazquez, and F. Casanova, Advanced Materials Interfaces **6**, 1900475 (2019).
[12]	J. Wunderlich, B. Kaestner, J. Sinova, and T. Jungwirth, Physical Review Letters **94**, 047204, 047204 (2005).
[13]	M. Isasa, E. Villamor, L. E. Hueso, M. Gradhand, and F. Casanova, Physical Review B **91**, 024402, 024402 (2015).
[14]	D. Qu, S. Y. Huang, B. F. Miao, S. X. Huang, and C. L. Chien, Physical Review B **89**, 140407, 140407 (2014).
[15]	L. Q. Liu, T. Moriyama, D. C. Ralph, and R. A. Buhrman, Physical Review Letters **106**, 036601, 036601 (2011).
[16]	Y. M. Lu, Y. Choi, C. M. Ortega, X. M. Cheng, J. W. Cai, S. Y. Huang, L. Sun, and C. L. Chien, Physical Review Letters **110**, 147207, 147207 (2013).
[17]	S. Geprags, S. Meyer, S. Altmannshofer, M. Opel, F. Wilhelm, A. Rogalev, R. Gross, and S. T. B. Goennenwein, Applied Physics Letters **101**, 262407, 262407 (2012).
[18]	N. Atodiresei and K. V. Raman, Mrs Bulletin **39**, 596 (2014).
[19]	O. Cespedes, M. S. Ferreira, S. Sanvito, M. Kociak, and J. M. D. Coey, Journal of Physics-Condensed Matter **16**, L155, Pii s0953-8984(04)76015-4 (2004).
[20]	M. Cinchetti, V. A. Dediu, and L. E. Hueso, Nature Materials **16**, 507 (2017).
[21]	K. Bairagi *et al.*, Physical Review Letters **114**, 247203 (2015).
[22]	F. A. Ma'Mari *et al.*, Nature **524**, 69 (2015).
[23]	F. Al Ma'Mari *et al.*, Proceedings of the National Academy of Sciences of the United States of America **114**, 5583 (2017).
[24]	F. Djeghloul *et al.*, The Journal of Physical Chemistry Letters **7**, 2310 (2016).
[25]	K. V. Raman *et al.*, Nature **493**, 509 (2013).
[26]	T. Moorsom *et al.*, Physical Review B **90**, 125311 (2014).
[27]	T. Moorsom *et al.*, Physical Review B **101**, 060408 (2020).
[28]	D. Sun, K. J. van Schooten, M. Kavand, H. Malissa, C. Zhang, M. Groesbeck, C. Boehme, and Z. V. Vardeny, Nature Materials **15**, 863 (2016).
[29]	R. Das *et al.*, Aip Advances **8**, 055906 (2018).
[30]	M. C. Wheeler *et al.*, Nature Communications **8**, 926, 926 (2017).
[31]	See Supplemental Material at http://link.aps.org/ for growth and microscopy methods, additional experimental results and further details on DFT simulations.
[32]	M. Grosser and U. Schmid, Thin Solid Films **517**, 4493 (2009).
[33]	S. Velez *et al.*, Physical Review Letters **116**, 016603 (2016).

[34]  S. R. Marmion, M. Ali, M. McLaren, D. A. Williams, and B. J. Hickey, Physical Review B **89**, 220404, 220404 (2014).
[35]  S. Meyer, M. Althammer, S. Geprags, M. Opel, R. Gross, and S. T. B. Goennenwein, Applied Physics Letters **104**, 242411 (2014).
[36]  X. P. Zhang, F. S. Bergeret, and V. N. Golovach, Nano Letters **19**, 6330 (2019).
[37]  Y. Tserkovnyak, A. Brataas, and G. E. W. Bauer, Physical Review B **66**, 224403 (2002).
[38]  J. G. Choi, J. W. Lee, and B. G. Park, Physical Review B **96**, 174412, 174412 (2017).
[39]  R. Friedrich, V. Caciuc, G. Bihlmayer, N. Atodiresei, and S. Blugel, New Journal of Physics **19**, 043017 (2017).
[40]  M. Isasa *et al.*, Physical Review B **93**, 014420 (2016).
[41]  S. Jakobs *et al.*, Nano Letters **15**, 6022 (2015).
[42]  J. C. Rojas Sanchez, L. Vila, G. Desfonds, S. Gambarelli, J. P. Attane, J. M. De Teresa, C. Magen, and A. Fert, Nature Communications **4**, 2944 (2013).
[43]  A. Mitra *et al.*, Scientific Reports **7**, 11774, 11774 (2017).
[44]  H. Corte-Leon, V. Nabaei, A. Manzin, J. Fletcher, P. Krysteczko, H. W. Schumacher, and O. Kazakova, Scientific Reports **4**, 6045, 6045 (2014).
[45]  W. Gil, D. Gorlitz, M. Horisberger, and J. Kotzler, Physical Review B **72**, 134401, 134401 (2005).
[46]  S. Dushenko, M. Hokazono, K. Nakamura, Y. Ando, T. Shinjo, and M. Shiraishi, Nature Communications **9**, 3118, 3118 (2018).